\documentclass{ISPIV2025_Template_Paper_Class}
\usepackage{booktabs}
\usepackage{multirow}

\begin{document}
\title{On Uncertainty Prediction for Deep-Learning-based Particle Image Velocimetry}
\author{ Wei Wang$^{1}$, Jeremiah Hu$^{1}$, Jia Ai$^{1}$, Yong Lee$^{1*}$}
\affiliation{
$^{1}$ Hubei Provincial Engineering Research Center of Robotics \& Intelligent Manufacturing, School of Mechanical and Electronic Engineering, Wuhan University of Technology, Wuhan 430070, China
        %
\vspace{7pt}$^*$ yonglee@whut.edu.cn
}
\maketitle



\begin{abstract}
Particle Image Velocimetry (PIV) is a widely used technique for flow measurement that traditionally relies on cross-correlation to track the displacement. Recent advances in deep learning-based methods have significantly improved the accuracy and efficiency of PIV measurements.
However, despite its importance, reliable uncertainty quantification for deep learning-based PIV remains a critical and largely overlooked challenge.
This paper explores three methods for quantifying uncertainty in deep learning-based PIV: the Uncertainty neural network (UNN), Multiple models (MM), and Multiple transforms (MT). 
We evaluate the three methods across multiple datasets. The results show that all three methods perform well under mild perturbations. Among the three evaluation metrics, the UNN method consistently achieves the best performance, providing accurate uncertainty estimates and demonstrating strong potential for uncertainty quantification in deep learning-based PIV.
This study provides a comprehensive framework for uncertainty quantification in PIV, offering insights for future research and practical implementation.
\end{abstract}

\section{Introduction}
Particle image velocimetry (PIV) is a popular technique for non-invasive flow measurement that relies on cross-correlation analysis to estimate the displacements~\cite{adrian1984scattering,raffel2018particle,lee2021diffeomorphic,ai2025rethinking}. While traditional methods provide velocity estimation through cross-correlation, they struggle with sparse resolution and computational efficiency. Recently, deep learning models such as PIV-DCNN~\cite{lee2017piv}, RAFT-PIV~\cite{lagemann2021deep}, and PIV-FlowDiffuser~\cite{zhu2025piv} have made significant progress by leveraging the power of deep learning to improve the accuracy and efficiency of PIV measurements. Using the robustness and adaptability of deep learning algorithms, these methods have significantly improved the accuracy and efficiency of PIV measurements.
However, these deep learning-based models intrinsically lack uncertainty estimation~\cite{sciacchitano2019uncertainty}, a critical drawback for applications requiring reliability, such as turbulent flow analysis or clinical hemodynamics. 
As a result, the limited ability of current deep neural network methods to quantify uncertainty hinders their adoption in practical PIV experiments.

Among the uncertainty quantification methods, existing quantification methods derive uncertainty measures from predetermined features, such as theoretical modeling of measurement chains, a posterior quantification methods such as uncertainty surfaces\cite{sciacchitano2019uncertainty}, and existing hand-crafted methods such as correlation statistics~\cite{wieneke2015piv,lee2024surrogate}, correlation moments~\cite{bhattacharya2018particle}, mutual information~\cite{xue2015particle}, etc. Therefore, they are not suitable for uncertainty estimation of neural networks in PIV applications. Meanwhile, deep learning models can construct complex mappings by learning from training data, providing a new idea that directly predict the uncertainty with artificial neural networks.
In the Bayesian convolutional neural network (BCNN)~\cite{morrell2021particle}, multiple samples generated from the output of the network or uncertainty parameters are incorporated into the model to obtain a more accurate estimate of the uncertainty associated with the velocity field, that is, the average and standard deviation calculated from $2000$ neural network outputs are used to quantify the uncertainty simultaneously, showing good performance, but at a high computational cost. Despite these advances, deep learning-based PIV methods still face challenges in uncertainty estimation and require further improvement.

Given the uncertainty quantification methods in other deep learning-based applications~\cite{abdar2021review}, Monte Carlo methods (multiple neural networks), Bootstrap methods (multiple input data), or probabilistic models (another uncertainty network) have not been explored in deep learning based PIV. 
Consequently, we explore the use of an Uncertainty Neural Network (UNN)—a variant of convolutional neural network-based measurement methods—along with two sampling-based approaches: the Multiple Models (MM) method and the Multiple Transforms (MT) method, to estimate the velocity uncertainty in the PIV field.
Specifically, we use the classic optical flow-based RAFT model~\cite{lagemann2021deep}, which is specially designed for PIV problems, to calculate the PIV velocity field, and use the velocity results as the input of the uncertainty estimation method. The velocity uncertainty is obtained through three methods. We verify and evaluate the three proposed methods on a PIV dataset (CAI for short)~\cite{cai2019dense}. Using $95\%$ coverage~\cite{timmins2012method}, Spearman's rank correlation coefficient (CC)~\cite{wannenwetsch2017probflow} and sparsification plots and area under curve (AUC) between predicted uncertainty and error~\cite{wannenwetsch2017probflow}, the UNN method performed best in measuring uncertainty results, followed by MM, and the performance of MT was relatively weak. The MM and MT methods demonstrate better generalization performance.
In total, this work addresses the lack of uncertainty prediction for deep learning-based PIV. The statistical uncertainty could enable potential applications for deep learning-based PIV.

\section{Uncertainty prediction methods}

This section introduces three methods for calculating the uncertainty of particle image velocimetry in detail, namely UNN (uncertainty neural network), MM (multiple models), and MT (multiple transforms). It explains the specific implementation principles, advantages and disadvantages, as well as the training dataset of each method. Evaluation indicators are also given.

\subsection{Uncertainty with uncertainty neural network}
\begin{figure}[ht]
    \centering
    \includegraphics[width=0.75\textwidth]{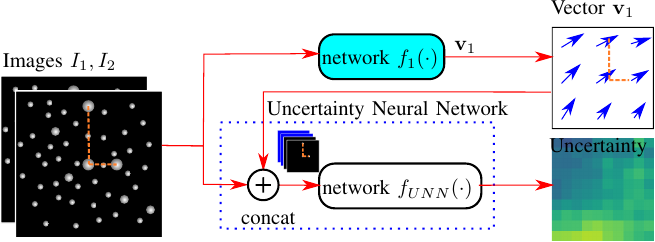} 
    \caption{The process of UNN method. Our UNN  predicts the measurement uncertainty via a neural network. Note that, the estimated velocity and the particle image pair are fed into the neural network.}
    \label{fig:1}
\end{figure}

The UNN (Uncertainty Neural Network) is an end-to-end neural network to estimate velocity uncertainty in particle image velocity measurement, details shown in Figure.~\ref{fig:1}. We first use the pre-trained RAFT neural network with the flow images in the training dataset to obtain the predicted velocity field. Then we try a U-net network inputted with the predicted velocity and flow images together. In the U-net neural network, the encoder extracts the multichannel features of flow images for regression analysis to obtain the velocity uncertainty. The network is trained with the following loss function~(Eq.~\ref{Train Loss}),
\begin{equation}
    \label{Train Loss}
    \mathcal{L} = -\frac{1}{N} \sum_{i=1}^{N} \left( \log \sigma_i + \frac{(e_i)^2}{2\sigma_i^2} \right)
\end{equation}
where $\sigma_i$ is the uncertainty estimation of UNN, $e_i = v_{gflow}- v_{flow}$ is the error between ground flow and predicted flow and $N$ is batch size. Compared to traditional methods, it can avoid the complex post-processing process (such as multiple iterations or manual feature extraction) and improve computational efficiency. The U-net neural network can focus on the regions with large prediction errors, such as flow field boundaries and occlusion areas, due to its multi-scale processing for images.
\subsection{Uncertainty with multiple models}
\begin{figure}[ht]
    \centering
    \includegraphics[width=0.75\textwidth]{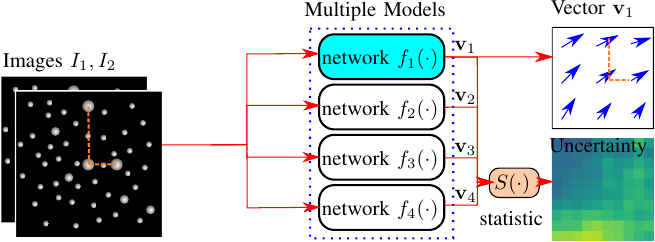} 
    \caption{The process of multiple models method. We use several different neural networks to deal with the flow images to get the same number velocity and then calculate the uncertainty.}
    \label{fig:2}
\end{figure}
MM (multiple models) is a method to calculate velocity uncertainty based on multi-model weight prediction and statistics. As shown in Figure~\ref{fig:2}, we use the multiple neural network instances $f_i$ to estimate the velocity respectively on the particle images $I_1$ and $I_2$, and then calculate uncertainty statistically. In this method, the four RAFT models are used, and the uncertainty of velocity is averaged.

This repeated measurement strategy effectively ensures the accuracy of uncertainty through multiple model predictions. While increasing the number of models enhances uncertainty estimation accuracy, it requires training multiple model instances and incurs higher computational costs.
\subsection{Uncertainty with multiple transforms}
\begin{figure}[htbp]
    \centering
    \includegraphics[width=0.75\textwidth]{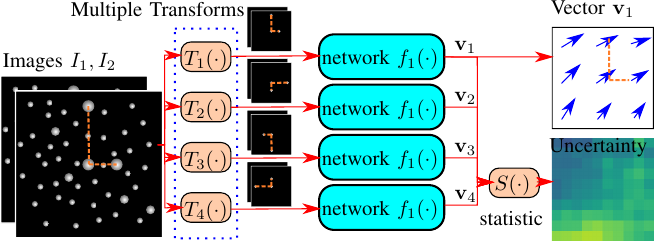} 
    \caption{The process of multiple transforms method. The flow images which are rotated to different angels are used to predict comparable velocity through the same neural network. The uncertainty is calculated by the velocity which are transformed back to right angles.}
    \label{fig:3}
\end{figure}
The MT (multiple transforms) is an image enhancement method such as rotation transformation. The input images are processed multiple times to obtain multiple sampling targets, which are inputted by the neural network to estimate velocity uncertainty. As shown in Figure~\ref{fig:3}, we transform four angles ($0^\circ, 90^\circ, 180^\circ, 270^\circ$) of the optical flow images and sent them into RAFT neural network to obtain four velocity field. Also, the obtained velocity fields are in four directions so that we transform them back to get right flow fields and then calculate uncertainty statistically.

Compared with the MM method, the difference of MT is the use of random sampling and statistical analysis based on the Monte Carlo method. After the image enhancement operation, the predicted velocity uncertainty has a certain degree of robustness, and can show good anti-interference ability.

The RAFT neural network is used as the core estimation method for particle image velocity field estimation. The model is trained and evaluated on the CAI dataset, which contains diverse flow field patterns including Uniform, JHTDB\_channel, DNS\_turb, SQG, backstep and Cylinder. The dataset is partitioned in a $7\mathbin{:}3$ ratio for training and testing respectively, ensuring stratified sampling across all flow types to maintain representative distributions in both sets. All experiments are conducted on an NVIDIA RTX $3090$ GPU platform, with the estimated flow fields subsequently serving as input to the UNN neural network for downstream processing.
\section{Experiment}
\subsection{Evaluation criteria}
In the deep learning-based PIV, the accuracy of the velocity field depends on the complex mapping learned from the training data. And we can ensure it through improving the uncertainty of the velocity field.
To evaluate the uncertainty estimation methods, we adopt three criteria, namely $95$\% coverage, Spearman's rank correlation coefficient (CC), and area under curve (AUC).

\subsubsection{95\% Coverage}
Coverage rate is used to evaluate the predicted uncertainty~\cite{timmins2012method}. Specifically, it is the proportion of data points that the absolute value of the error is less than the absolute value of the uncertainty, as shown in the following~(Eq.~\ref{Coverage}).
\begin{equation}
    \label{Coverage}
    \text{Coverage}_{95\%} = \frac{1}{N} \sum_{i=1}^{N} \mathbf{1} \left( |e_i| < \sigma_i \right)
\end{equation}
where $\mathbf{1}(\cdot)$ means if the condition in the brackets is true, the value is $1$, otherwise the value is 0. According to the $2\sigma$ principle of the Gaussian distribution, the expected coverage rate is about $95$\%. If the actual coverage rate is closer to the theoretical value,  we will get more accurate uncertainty estimation. By calculating the error of $U$ and $V$ respectively, we determine that the predicted value falls within twice range of the uncertainty prediction value.
\subsubsection{Correlation coefficient}
The Spearman's rank correlation coefficient~\cite{wannenwetsch2017probflow} is used to calculate the monotonic correlation and quantify the statistical correlation between the predicted uncertainty and the absolute value of the error, the specific calculation formula being the following~~(Eq.~\ref{CC}):
\begin{equation}
    \label{CC}
    \rho = 1 - \frac{6 \sum_{i=1}^{N} (r_{e_i} - r_{\sigma_i})^2}{N(N^2 - 1)}
\end{equation}
where $r_{e_i}$ and $r_{\sigma_i}$ are rank of $e_i$ and $\sigma_i$separately.
The metric ranges from $-$1 to 1 where 1($-$1) denotes perfect positive (negative) correlation. Values approaching $1$ imply larger errors in high uncertainty regions, confirming the reliability of uncertainty estimation. We use the EPE endpoint error as the true value, and use the product of the uncertainty of $u$ and $v$ estimated by the three methods to calculate the sparsity curve and visualize it, which intuitively reflects the monotonic correlation.
\subsubsection{Area under curve}
Area under curve~\cite{wannenwetsch2017probflow} is based on the sparsification curve to calculate the curve area under the corresponding data points~(Eq.~\ref{AUC}). 
\begin{equation}
    \label{AUC}
    \text{AUC} = \frac{1}{N} \sum_{i=1}^{N} \frac{1}{\bar{e}} \sum_{j=1}^{i} e_j
\end{equation}
\begin{equation}
    \bar{e} = \frac{1}{N} \sum_{i=1}^{N} e_i
\end{equation}
where $e_j$ is the uncertainty errors sorted from smallest to largest and $\bar{e}$ is the mean error.
The low uncertainty points are gradually eliminated with the normalized average error of the remaining points calculated. A lower AUC indicates higher uncertainty and larger errors, with reduced residual errors confirming that uncertainty classification reliably predicts the magnitude of the error.
To evaluate the performance of methods, we conduct experiments on the divided test datasets, some synthetic datasets, and natural datasets according to the above evaluation criteria, and obtain relevant results.

\subsection{Results of CAI test images}
\begin{figure}[ht]
\centering
\includegraphics[width=1.0\textwidth,height=0.8\textwidth]{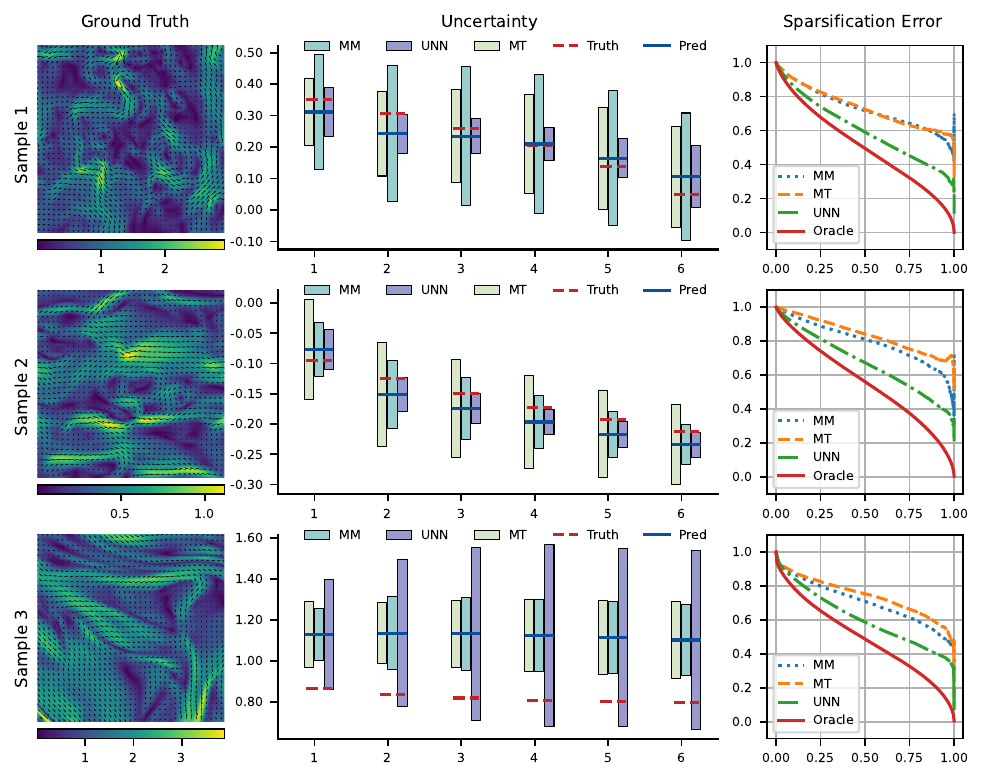}
    \caption{The performance of three methods on sample data.}
    \label{fig:4}
\end{figure}

The UNN neural network trained with the CAI dataset is evaluated on the divided test dataset and compared with the other two methods. We sampled three from the test dataset to visualize the evaluation results shown in Figure.~\ref{fig:4}. The central part displays the predicted velocity uncertainty and its associated error bounds for the three methods, evaluating both the agreement between predictions and ground truth, as well as the coverage of the true values within the estimated uncertainty ranges. The rightmost plot illustrates the correlation between the predicted uncertainty and the actual error magnitude.
\begin{table}[h]
    \centering
    \vspace{-15pt}
    \caption{The performance of CAI test images measured with $95$\% Coverage, AUC and CC. Best in \textbf{bold}.}
    \label{tab:combined-results}
    \begin{tabular}{@{}cccccccccc@{}}
    \toprule
    \multirow{2}{*}{Method} & \multicolumn{3}{c}{Sample 1} & \multicolumn{3}{c}{Sample 2} & \multicolumn{3}{c}{Sample 3}\\ 
    \cmidrule(lr){2-4} \cmidrule(lr){5-7} \cmidrule(lr){8-10} 
                            & $95$\%Coverage & AUC & CC & $95$\%Coverage & AUC & CC & $95$\%Coverage & AUC & CC \\  
    \midrule 
            MM              & 0.67 & 0.73 & 0.49 & 0.66& 0.80& 0.40& 0.75& 0.70& 0.46\\
            MT              & 0.75 & 0.74 & 0.47 & 0.78& 0.84& 0.32& 0.82& 0.74& 0.40\\
            UNN             & \textbf{0.86} & \textbf{0.60} & \textbf{0.77}& \textbf{0.88}& \textbf{0.67}& \textbf{0.72}& \textbf{0.83}& \textbf{0.61}& \textbf{0.73}\\ 
    \bottomrule 
    \\[-1.5em] 
    \end{tabular}
\end{table}

The confidence interval obtained by our UNN method effectively covers the truth value and visually outperforms those of MM and MT. The sparsification plots reveal that the UNN method can accurately reflect the actual errors, because the UNN is closer to the ideal uncertainty (Oracle). Besides, the performance advantages of the UNN method are also demonstrated in the metrics presented in Table.~\ref{tab:combined-results}.

\subsection{Test with Gaussian noise and blur}
Since PIV images have natural noise, we apply two perturbations, Gaussian noise and Gaussian blur, to the optical flow images to simulate the image degradation caused by sensor noise or motion blur in actual scenes, and analyze the impact of the perturbations on the three prediction methods.

We add Gaussian noise with mean value $0$ and different variance $var$, which takes values of $0$, $5$ and $10$ on the test images of dataset CAI. Similarly, the Gaussian blur kernel with standard deviation $\sigma$ (blur intensity) which takes values of $0$, $2.5$, and $5$ is added to the same flow images. These parameter values create a progressive degradation scale where higher values correspond to stronger corruption intensities. By testing across this gradient of distortion levels, we systematically examine how different noise magnitudes and blur intensities affect the model's performance.
\begin{table}[h]
    \centering
    \vspace{-15pt}
    \caption{The performance of \textbf{Gaussian noise} measured with $95$\%Coverage, AUC and CC from all CAI test images. Best in \textbf{bold}.}
    \label{tab:gaussian noise}
    \begin{tabular}{@{}cccccccccc@{}}
    \toprule
    \multirow{2}{*}{Method} & \multicolumn{3}{c}{$var$ = 0} & \multicolumn{3}{c}{$var$ = 5} & \multicolumn{3}{c}{$var$ = 10}\\ 
    \cmidrule(lr){2-4} \cmidrule(lr){5-7} \cmidrule(lr){8-10} 
                            & $95$\%Coverage & AUC & CC & $95$\%Coverage & AUC & CC & $95$\%Coverage & AUC & CC \\  
    \midrule 
            MM              & 0.67 & 0.73 & 0.49 & 0.68& 0.75& 0.49& 0.70& 0.80& 0.37\\
            MT              & 0.77 & 0.74 & 0.47 & 0.77& 0.75& 0.45& 0.79& \textbf{0.76}&\textbf{0.43}\\
            UNN             & \textbf{0.90} & \textbf{0.62} & \textbf{0.73}& \textbf{0.86}& \textbf{0.72}& \textbf{0.51}& \textbf{0.83}& 0.80& 0.28\\ 
    \bottomrule 
    \\[-1.5em] 
    \end{tabular}
\end{table}

\begin{table}[h]
    \centering
    \vspace{-15pt}
    \caption{The performance of \textbf{Gaussian blur} measured with $95$\%Coverage, AUC and CC calculated from all CAI test images. Best in \textbf{bold}.}
    \label{tab:gaussian blur}
    \begin{tabular}{@{}cccccccccc@{}}
    \toprule
    \multirow{2}{*}{Method} & \multicolumn{3}{c}{$\sigma$ = 0} & \multicolumn{3}{c}{$\sigma$ = 2.5} & \multicolumn{3}{c}{$\sigma$ = 5}\\ 
    \cmidrule(lr){2-4} \cmidrule(lr){5-7} \cmidrule(lr){8-10} 
                            & $95$\%Coverage & AUC & CC & $95$\%Coverage & AUC & CC & $95$\%Coverage & AUC & CC \\  
    \midrule 
            MM              & 1.00 & 0.93 & 0.82 & 1.00& \textbf{0.94}& \textbf{0.69}& 1.00& 1.08& -0.33\\
            MT              & 1.00 & 0.96 & 0.38 & 1.00& 0.96& 0.4& 1.00& 1.04&0.01\\
            UNN             & 1.00 & \textbf{0.92} & \textbf{0.88}& 0.99&0.95& 0.54& 0.98& \textbf{0.99}& \textbf{0.20}\\ 
    \bottomrule 
    \\[-1.5em] 
    \end{tabular}
\end{table}

The results are shown in the following Table.~\ref{tab:gaussian noise} and Table.~\ref{tab:gaussian blur}, the data in the tables being made statistics from all images.
Under low-intensity interference, the three methods can maintain accuracy, which the UNN method being the best. As the interference continues to increase, the evaluation performance of the three methods gradually decreases. When the disturbance reaches a high level, due to excessive image distortion, the accurate uncertainty cannot be obtained. The three methods fail at large distribution, and the evaluation results are in line with expectations.
\subsection{Performance on unseen images}
To evaluate the practical applicability of the uncertainty method, we test on the Sintel dataset and rotated flow field images. 
Sintel is a dataset for optical flow evaluation, with $1064$ synthetic stereo images and ground truth data for disparity. Stereo images are RGB and disparity is grayscale. The resolution of both is $1024\times436$ pixels, $8$ bits per channel. It is widely used in the evaluation of deep neural network model architectures in the field of PIV. In this experiment, we crop the center of the optical flow image in Sintel to $256\times256$ pixels, calculate the uncertainty predicted by the first two frames of optical flow images in $23$ scenes. According to the other dataset, the flow file rotates with a velocity of about $33.3$ rpm. We sampled $9$ pictures of $400\times400$ pixels, and used $2$ adjacent pictures as input to test. 
\begin{figure}[ht]
    \centering

    \begin{minipage}[c]{0.3\linewidth}
        \centering
        \includegraphics[width=\linewidth]{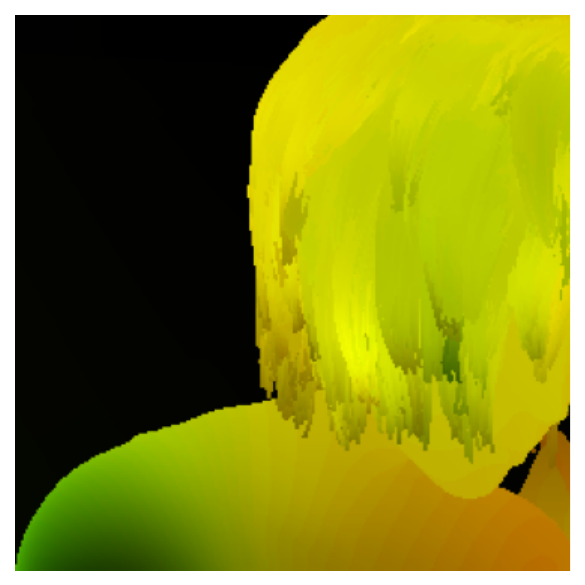} \\
        \hspace{-0.6cm}
        \includegraphics[width=\linewidth]{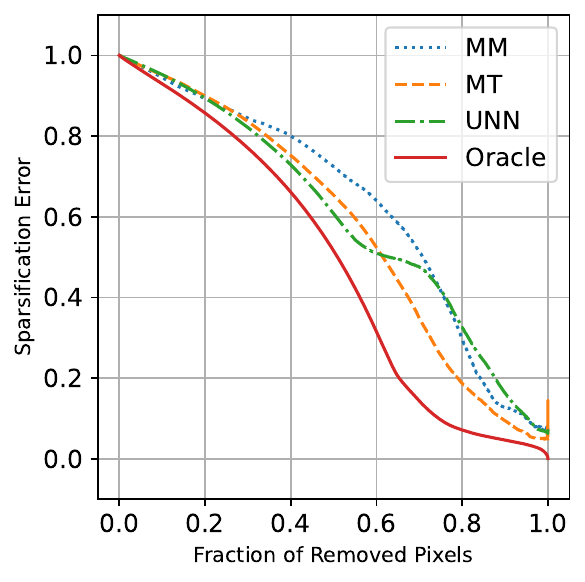} \\
        \small (a) Sintel: alley\_1
    \end{minipage}
    \hfill
    \begin{minipage}[c]{0.3\linewidth}
        \centering
        \includegraphics[width=\linewidth]{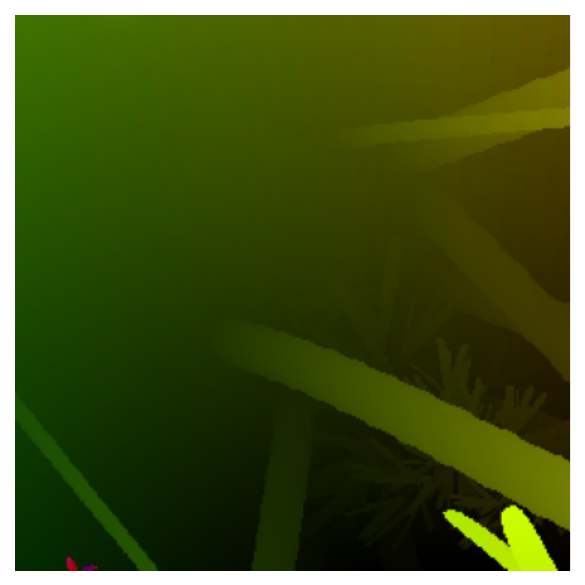} \\
        \hspace{-0.6cm}
        \includegraphics[width=\linewidth]{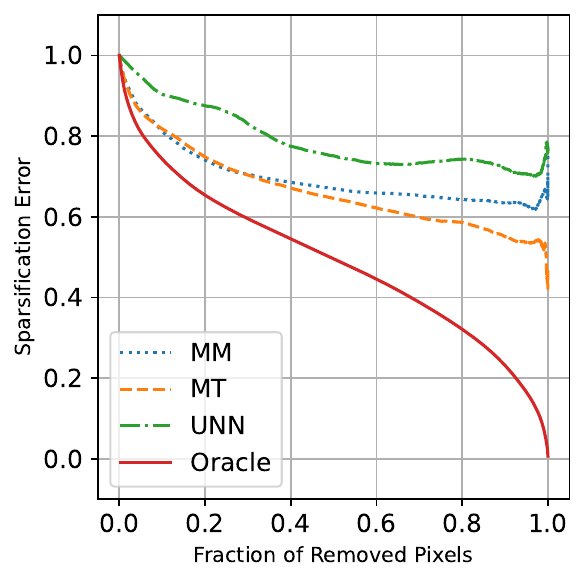} \\
        \small (b) Sintel: bamboo\_1
    \end{minipage}
    \hfill
    \begin{minipage}[c]{0.3\linewidth}
        \centering
        \includegraphics[width=\linewidth]{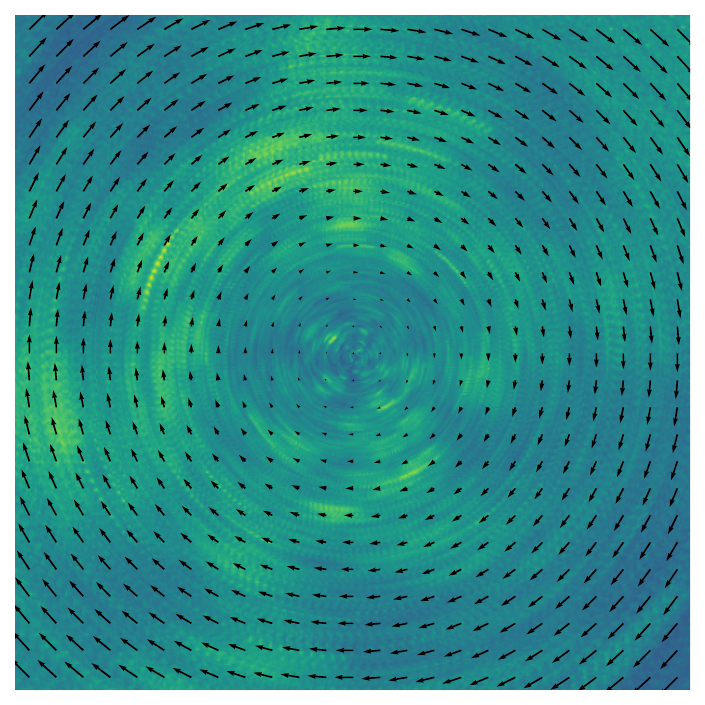} \\
        \hspace{-0.6cm}
        \includegraphics[width=\linewidth]{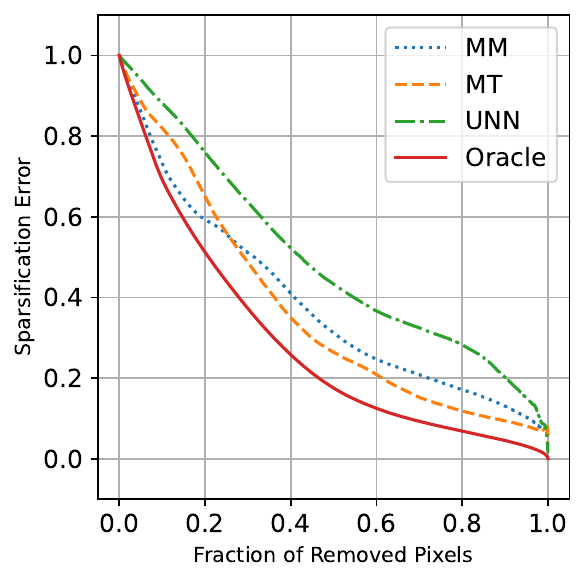} \\
        \small (c) Rotating flow
    \end{minipage}

    \caption{The performance of unseen images in three groups, each group is consist of a ground flow image and a Sparsification error graph. Group (a) and group (b) are two samples of Sintel test images and group (c) shows performance of rotating flow filed.}
    \label{fig:grouped_images}
\end{figure}
\begin{table}[h]
    \centering
    \vspace{-15pt}
    \caption{The performance of unseen images measured with $95$\%Coverage, AUC and CC. Best in \textbf{bold}.}
    \label{tab:unseen images}
    \begin{tabular}{@{}cccccccccc@{}}
    \toprule
    \multirow{2}{*}{Method} & \multicolumn{3}{c}{Sintel: alley\_1} & \multicolumn{3}{c}{Sintel: bamboo\_1} & \multicolumn{3}{c}{Rotating flow}\\ 
    \cmidrule(lr){2-4} \cmidrule(lr){5-7} \cmidrule(lr){8-10} 
                            & $95$\%Coverage & AUC & CC & $95$\%Coverage & AUC & CC & $95$\%Coverage & AUC & CC \\  
    \midrule 
            MM              & \textbf{0.57} & 0.63 & 0.56 & \textbf{0.67} & 0.69 & 0.42 & 0.79 & 0.38 & 0.76\\
            MT              & 0.49 & \textbf{0.57} & 0.65 & 0.63 & \textbf{0.67} & \textbf{0.53} & \textbf{0.84} & \textbf{0.36} & \textbf{0.82}\\
            UNN             & 0.05 & 0.60 & \textbf{0.66} & 0.12 & 0.79 & 0.23 & 0.26 & 0.49 & 0.61\\ 
    \bottomrule 
    \\[-1.5em] 
    \end{tabular}
\end{table}

We sample $2$ scenes of Sintel shown in the first and second columns of Figure.~\ref{fig:grouped_images}. The corresponding evaluation data are shown in Table.~\ref{tab:unseen images}. Our UNN neural network, which has not been trained with Sintel, performs poorly in terms of error coverage, lower than the sampling methods MM and MT. However, in some scenarios, the error correlation obtained by the UNN method is still in a preferred position, indicating that its error estimation is convincing. The MM and MT methods perform better, thanks to the stability of the traditional methods. The third column shows the result of the rotated optical flow field experiment. We can find that the MM and MT methods are better than the UNN method, but the difference is not large. In summary, the three methods have limited generalization capability on this dataset.

\section{Conclusion}
In this study, we test three methods to quantify the uncertainty in deep learning-based particle image velocimetry (PIV). And they are uncertainty neural network (UNN), multiple models (MM), and multiple transforms (MT). UNN is an end-to-end neural network that combines RAFT for velocity prediction and U-net for uncertainty regression, and performs well in capturing error-sensitive areas while maintaining computational efficiency. MM and MT, which exploit multi-model ensemble and image transformation, respectively, provide robust uncertainty estimates with better generalization capabilities, albeit at a higher computational cost. Our work provides a systematic framework for evaluating uncertainty in deep learning-based PIV on different images, combining $95$\% Coverage, AUC, and CC metrics. Together, these metrics verify the reliability of uncertainty estimates under different conditions. We will expand research into uncertainty quantification methodologies for deep learning-based PIV.
\begin{acknowledgments}
Supported by the Hubei Provincial Natural Science Foundation of China (Grant No. 2023AFB128). The authors thank Mr. Ding Pan for a beneficial discussion. The authors appreciate the anonymous reviewers for their careful comments and professional suggestions. This paper has been accepted for presentation at the 16th International Symposium on Particle Image Velocimetry (ISPIV 2025).
\end{acknowledgments}
\bibliography{ISPIV2025_References}

\end{document}